\documentclass[acmsmall,screen]{acmart}

\AtBeginDocument{%
  }
\usepackage[ruled,linesnumbered]{algorithm2e}
\usepackage{multirow}
\usepackage{amsmath}
\setcopyright{acmlicensed}
\copyrightyear{2018}
\acmYear{2018}
\acmDOI{XXXXXXX.XXXXXXX}

\acmJournal{JACM}
\acmVolume{37}
\acmNumber{4}
\acmArticle{111}
\acmMonth{8}

\begin{document}

\title{Backpropagation-Free Multi-modal On-Device Model Adaptation via Cloud-Device Collaboration}

\author{Wei Ji}
\authornote{These authors contributed equally to this research.}
\affiliation{%
  \institution{National University of Singapore}
  \country{Singapore}
}
\email{jiwei@nus.edu.sg}
\orcid{0000-0002-8106-9768}

\author{Li Li}
\authornotemark[1]
\affiliation{%
  \institution{National University of Singapore}
  \country{Singapore}
}
\email{lili02@u.nus.edu}

\author{Zheqi Lv}
\authornotemark[1]

\orcid{0000-0001-6529-8088}
\affiliation{%
  \institution{Zhejiang University}
  \country{China}}
\email{zheqilv@zju.edu.cn}

\author{Wenqiao Zhang}

\affiliation{%
  \institution{Zhejiang University}
  \country{China}
}
\email{wenqiaozhang@zju.edu.cn}

\author{Mengze Li}
\affiliation{%
 \institution{Zhejiang University}
 \country{China}
}
\email{mengzeli@zju.edu.cn} 

\author{Zhen Wan}
\affiliation{%
  \institution{Fudan University}
  \country{China}
  }
\email{wz2311602492@gmail.com}

\author{Wenqiang Lei}

\affiliation{%
  \institution{Sichuan University}
  \country{China}
}
\email{wenqianglei@gmail.com}
\orcid{0000-0001-6540-0601}

\author{Roger Zimmermann}

\affiliation{%
  \institution{National University of Singapore}
  \country{Singapore}}
  \email{dcsrz@nus.edu.sg}
  \orcid{0000-0002-7410-2590}

\renewcommand{\shortauthors}{Trovato et al.}

\begin{abstract}
In our increasingly interconnected world, where intelligent devices continually amass copious personalized multi-modal data, a pressing need arises to deliver high-quality, personalized device-aware services. However, this endeavor presents a multifaceted challenge to prevailing artificial intelligence (AI) systems primarily rooted in the cloud. As these systems grapple with shifting data distributions between the cloud and devices, the traditional approach of fine-tuning-based adaptation (FTA) exists the following issues: the costly and time-consuming data annotation required by FTA and the looming risk of model overfitting. To surmount these challenges, we introduce a Universal On-Device Multi-modal Model Adaptation Framework, revolutionizing on-device model adaptation by striking a balance between efficiency and effectiveness. The framework features the Fast Domain Adaptor (FDA) hosted in the cloud, providing tailored parameters for the Lightweight Multi-modal Model on devices. 
To enhance adaptability across multi-modal tasks, the AnchorFrame Distribution Reasoner (ADR) minimizes communication costs. Our contributions, encapsulated in the Cloud-Device Collaboration Multi-modal Parameter Generation (CDC-MMPG) framework, represent a pioneering solution for on-Device Multi-modal Model Adaptation (DMMA). Extensive experiments validate the efficiency and effectiveness of our method, particularly in video question answering and retrieval tasks, driving forward the integration of intelligent devices into our daily lives.
\end{abstract}

\ccsdesc[500]{Information systems~Mobile information processing systems}
\ccsdesc[500]{Information systems~Personalization}
\ccsdesc[500]{Human-centered computing~Mobile computing}

\keywords{Cloud-device collaboration, model adaptation, multi-modal}


\maketitle

\section{Introduction}
\begin{figure}[t]
  \centering
  \includegraphics[width=\textwidth]{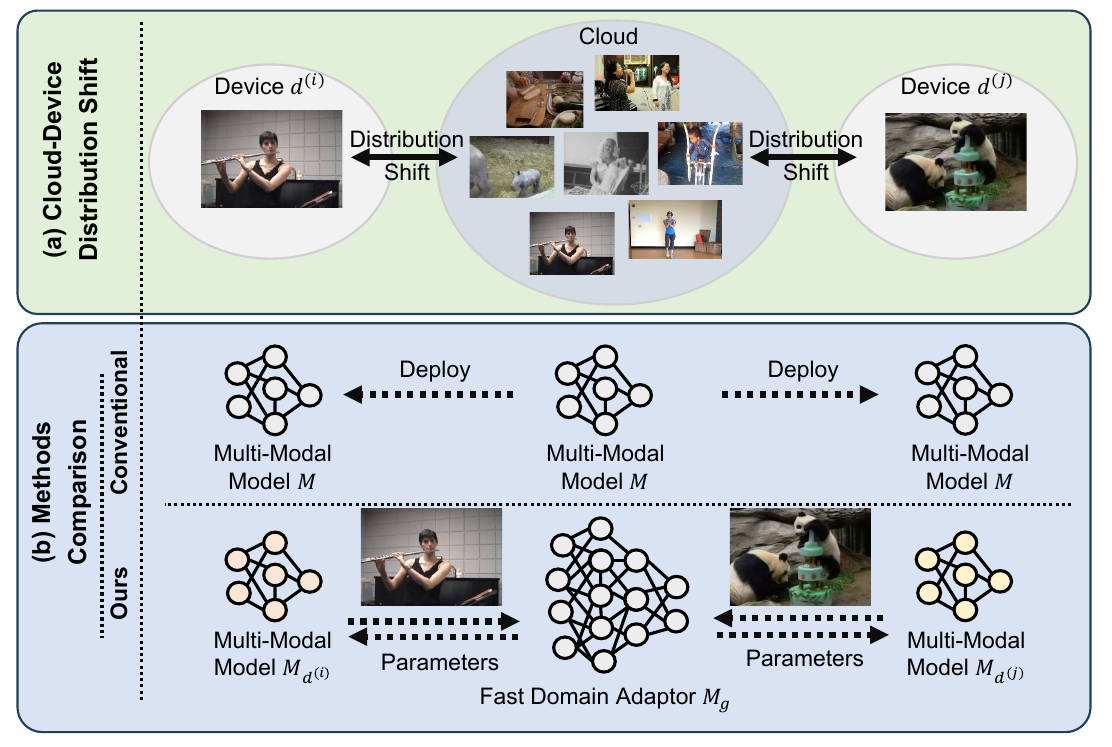}

  \caption{ (a) Multi-modal data on cloud and different devices exist in different distributions due to the personalized preference of users. (b) Compared with conventional methods of deploying models on different devices, we propose an FDA that can achieve a balance of efficiency and effectiveness.}\label{introduction}
\end{figure}

In today's interconnected world, the proliferation of intelligent devices, ranging from ubiquitous smartphones to the ever-expanding Internet of Things (IoT) ecosystem, has become an integral part of our daily lives. These devices serve as data collection powerhouses, continuously amassing vast repositories of personalized multi-modal data, which can include a wide array of input modalities such as text, images and videos.
The potential locked within this trove of multi-modal data arriving continuously is immense,  promising to unlock high-quality and tailored device-aware services for individual users. Despite promising, the personalized device service involves analyzing the dynamic nature of the multi-modal data that underscore users' intentions. The prevailing artificial intelligence (AI) systems, primarily trained and deployed in cloud-based environments, face a profound challenge in adapting to the dynamic device data when using a static cloud model for all individual users, mainly due to the distribution shift of the cloud and device data, as shown in Figure~\ref{introduction}.  In other words, high-quality personalized service requires AI systems to undergo continual refinement and adaptation to accommodate the evolving landscape of personalized multi-modal data.


 

Intuitively, one of the straightforward adaptation strategies is to fine-tune the cloud model based on the device's multi-modal data, which can kindly alleviate the cloud-device data distribution shift to model users' intentions.  Nevertheless, we contend that the fine-tuning-adaptation (FTA) paradigm 
may not satisfactorily resolve device model personalization, which can be summarized as two key aspects: (1) \textbf{Undesirable Annotation.} FTA  often necessitates manually annotating data to guide model adaptation, which typically hinges on expensive and labor-intensive device data labeling.  Additionally, this retraining process can result in substantial delays, hindering the AI system's ability to deliver real-time, context-aware responsiveness. This situation is further exacerbated by the inherent complexity of multi-modal data understanding, \emph{i.e.}, with data streams that encompass textual, visual, and auditory information, the intricacies of labeling and retraining are more time-consuming, resulting in a higher time delay and thus diminishing its practicality for users. (2) \textbf{Overfitting Risk.} In real-world applications, the majority of devices could be characterized by sparse and specialized multi-modal data where model fine-tuning may inadvertently lead to overfitting issues, even leading to device model performance degradation. In other words, FTA imposes significant demands for on-device data quantity, hampering its ability to generalize effectively across diverse device ecosystems. Based on the aforementioned insights, a meaningful optimization goal of device model personalization is to appropriately tap onto the personalized multi-modal data, and thus strike the delicate balance between the effectiveness and efficiency of personalized adaptation.


To address these multifaceted challenges and pave the way for intelligent device-driven AI systems that harmonize adaptability and computational efficiency, we present a universal device multi-modal model adaptation framework, a groundbreaking approach that redefines the landscape of on-device AI adaptation. This framework is designed to revolutionize the way AI systems harness multi-modal data, simultaneously unlocking efficiency and effectiveness in the adaptation process.
In our approach, we introduced a Fast Domain Adaptor (FDA) hosted on the cloud, which takes device-captured images as input and produces customized parameters for the Lightweight Multi-modal Model on the device. These customized parameters are tailored to different data distributions present on the device. Moreover, the FDA adapts to various multi-modal tasks by requiring distinct data inputs. However, certain multi-modal tasks, like Video QA, necessitate transmitting extensive data such as multiple frames or the entire video to FDA, incurring substantial communication costs and bandwidth requirements. To mitigate this challenge and enhance FDA's adaptability across different multi-modal tasks, we developed the AnchorFrame Distribution Reasoner (ADR). ADR standardizes the input for the FDA across various multi-modal tasks. For instance, in the case of Video QA, ADR selects the first frame of each video as the AnchorFrame. Using a combination of Variation AutoEncoder (VAE) and KeyFrame, ADR maps the AnchorFrame to a standard distribution through training. Once both FDA and ADR are trained, they are deployed on the cloud to deliver personalized model parameter services for lightweight multi-modal models on the device. The device's model can then upload the AnchorFrame to obtain personalized model parameters tailored to the data distribution of the current image or video input.
An important highlight of our framework is the absence of backpropagation during the domain adaptation process for the device model. This unique approach enables our framework to achieve rapid domain adaptation of the device model, resulting in exceptionally low latency.

Summing up, our contributions are summarized below:
\begin{enumerate}
    \item We propose a \textbf{C}loud-\textbf{D}evice \textbf{C}ollaboration \textbf{M}ulti-\textbf{M}odal \textbf{P}arameter \textbf{G}eneration (CDC-MMPG) framework to accomplish efficient on-\textbf{D}evice \textbf{M}ulti-modal \textbf{M}odel \textbf{A}daptation (DMMA). Its core is Fast Domain Adaptor (FDA) which can generate multi-modal equipment model parameters based on the data distribution of multi-modal data. To the best of our knowledge, we are the first ones to solve the DMMA. Our proposed CDC-MMPG framework is a universal one that can be adaptive to other modalities. 
    
    \item We propose the AnchorFrame Distribution Reasoner (ADR) to further reduce the communication cost of the CDC-MMPG. ADR successfully addresses the issue of high bandwidth dependence while maintaining nearly consistent performance.
    
 
    \item Extensive experiments on video question answering and video retrieval tasks have verified the efficiency and effectiveness of our proposed method.
\end{enumerate}

\section{Related Works}

\textbf{Cloud-Device Collaboration.} In the rapidly evolving sphere of cloud-device collaboration, deep learning has become a crucial component, strategically combining the features of cloud-based and on-device machine learning. Although Federated Learning (FL), notably represented by FedAVG~\cite{collins2022fedavg}, has traditionally been a significant player, its perceived simplicity has limited its applicability in various practical situations. This dynamic field has seen the rise of innovative methods and techniques designed to optimize the interaction between cloud resources and mobile devices. For instance, MPDA~\cite{yan2022device} skillfully utilizes cloud-based samples to enhance the performance of on-device models. 
Another interesting approach involves deploying multiple models with similar functionalities, coordinated by a Meta Controller for optimal task allocation, thus expanding the possibilities of collaborative model adaptation \cite{zhang2022learningfreeobjectsegments, chen2023federatedlearningshareablebases}.  
Moreover, DUET~\cite{lv2023duet}, inspired by the HyperNetwork concept \cite{hyper3,xian2021hyperdynamics,zhang2020graph,oscar}, simplifies the device model adaptation process, eliminating the need for on-device training \cite{zhong2024makingbatchnormalizationgreat}. IDEAL~\cite{lv2023ideal} expands parameter generation-based models to the domain of recommender systems, particularly DC-CDR~\cite{zhang2023disentangled}. This enhances the generalization capabilities of device recommendation models, though it requires consideration of factors such as request frequency and communication revenue. However, the above works are tailored for recommender system, which can not be directly migrated to the multi-modal domain field, mainly due to that they ignore the inherent complexity of multi-modal data. In this paper, we present CDC-MMPG, tailored for multi-modal on-device model adaptation via cloud-device collaboration, which simultaneously obtained effectiveness and efficiency.

\textbf{Domain Adaptation.}
Domain Adaptation (DA) plays a crucial role in transferring a network that was initially trained on a labeled source domain to effectively perform on a target domain. 
Various techniques have been developed to align the feature distributions between the source and target domains, such as maximum mean discrepancy and correlation alignment. These methods aim to reduce the distribution discrepancy and enable the network to generalize well across different domains, improving its performance in the target domain. 
Lots of methods are proposed to facilitate the development of the domain adaptation, such as multi-stage methods~\cite{zhang2019curriculum, 
chen2021unsupervised,liicassp,jiweiacmtrans,Limm}, gradual transfer strategies~\cite{chen2019progressive,zhensuicse,li2024visionlanguagemodelfinetuningsimple,tu2023holistictransfernondisruptivefinetuning}, and so on. 
Recent advancements in domain adaptation embrace curriculum-based strategies inspired by curriculum learning to enhance the effectiveness of the adaptation process \cite{shu2019transferable, roy2021curriculum,Li_Ji_Wu_Li_Qin_Wei_Zimmermann_2024}, 
Source-free domain adaptation (SFDA) has emerged as a response to privacy and copyright concerns, where access to the source domain data is restricted or unavailable~\cite{zhang2019curriculum,chen2021unsupervised,shu2019transferable,chen2019progressive,roy2021curriculum,luo2020adversarial,10.1145/3528223.3530164,Ding_2022_CVPR,Wang_2022_CVPR,NEURIPS2022_26300457,Sun_2022_CVPR,Litrico_2023_CVPR}. SFDA presents a more challenging variant known as Test-Time Adaptation (TTA), which necessitates online adaptation during the inference phase.
In addition, many researchers introduce 
 the domain adaptation setting into different tasks to enhance model generalization \cite{ wang2023dance, mosallanezhad2022domain,Ahmed_2023_ICCV,da_Costa_2022_WACV,Yang_2023_ICCV,Ma_2022_CVPR,9695359,Zhang_2022_CVPR,9950721,9785619,9927350,10234713,NEURIPS2023_407106f4}.
In summary, domain adaptation (DA) methods focus on aligning feature distributions between the source and target domains. Recent advancements have highlighted the importance of curriculum-based strategies to improve the adaptation process.~\cite{zhang2019curriculum,chen2021unsupervised,shu2019transferable,chen2019progressive,roy2021curriculum,Ye_2022_CVPR}. In this paper, we propose a model generation framework to solve the personalized adaption problem.


\begin{figure*}
\includegraphics[width=1.0\textwidth]{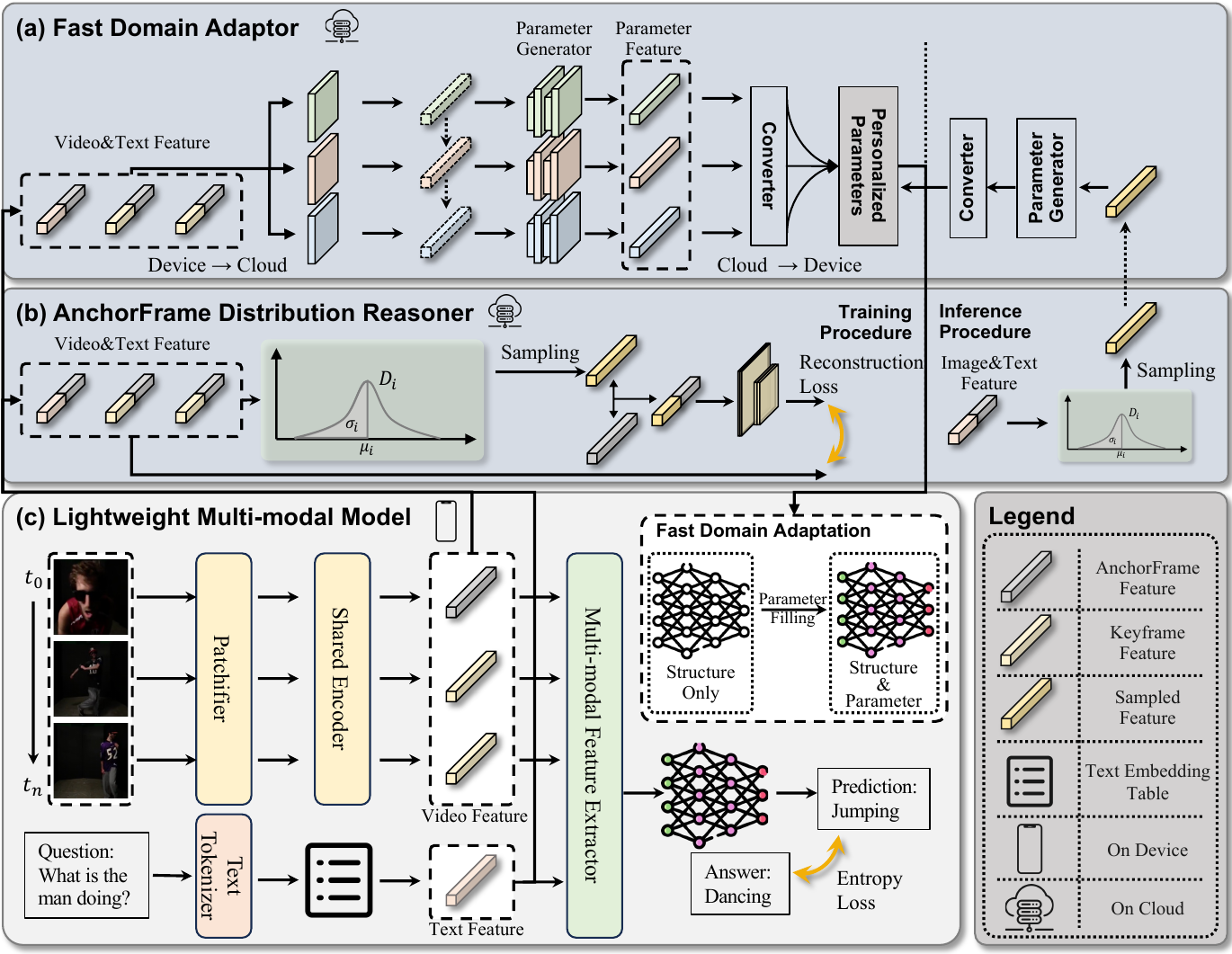} 

 \caption{Illustration of the overall pipeline of our method, CDC-MMPG. (a) and (b) represent the Cloud model, which reconstructs the video features uploaded from the device and reasons out the personal parameters of the device model based on the reconstructed video features. (c) represents the lightweight multi-modal device-side model, which extracts the multi-modal features, and uploads the video features to the cloud model for the personal device-model parameter prediction. After being updated with the personal parameters, the lightweight multi-modal device-side model will further analyze the multi-modal features and make the final prediction. }
 \label{img:model}
\end{figure*}

\section{Methodology}
In this section, we provide a detailed description of our proposed \textbf{C}loud-\textbf{D}evice \textbf{C}ollaboration \textbf{M}ulti-\textbf{M}odal \textbf{P}arameter \textbf{G}eneration (CDC-MMPG) framework, which includes \textbf{F}ast \textbf{D}omain \textbf{A}daptor (FDA) and \textbf{A}nchorFrame \textbf{D}istribution \textbf{R}easoner (ADR).


\subsection{Preliminary}

\textbf{Problem Formulation.}
For the on-Device Multi-modal Model Adaptation (DMMA) in the device-cloud collaboration system, we have access to a set of devices $\mathcal{D}=\{d^{(i)}\}_{i=1}^{\mathcal{N}_d}$, each device with its personal i.i.d multi-modal history samples  $\mathcal{S}_{H^{(i)}}$ and multi-modal real-time samples $\mathcal{S}_{R^{(i)}}$ in current session, where $\mathcal{N}_{d}$ represents the number of devices.
The goal of on-Device Multi-modal Model Adaptation is to generalize a trained cloud model $\mathcal{M}_g(\cdot;\Theta_g)$ learned from $\{\mathcal{S}_{H^{(i)}}\}_{j=1}^{\mathcal{N}_d}$ to each specific local device model $\mathcal{M}_{d^{(i)}}(\cdot;\Theta_{d^{(i)}})$ 
conditioned on real-time samples $\mathcal{S}_{R^{(i)}}$, where $\Theta_g$ and $\Theta_{d^{(i)}}$ respectively denote the learned parameters for the cloud model and the $i$-th device model:
\begin{equation}
\begin{aligned}
 \underbrace{\rm {\textbf{CDC-MMPG}}}_{\rm{DMMA\ Model}}: \underbrace{\mathcal{M}_{g}(\{\mathcal{S}_{H^{(i)}}\}_{i=1}^{\mathcal{N}_d};\Theta_g)}_{\rm{Cloud\ Model}}  \rightarrow
\underbrace{\mathcal{M}_{d^{(i)}}(\mathcal{S}_{R^{(i)}};\Theta_{d^{(i)}})}_{\rm{Device\ Model}}.
 \label{equ_1}
\end{aligned}
\end{equation}

Figure~\ref{img:model} illustrates the overview of our CDC-MMPG framework which consists of two modules to improve the generalization ability of the trained models on the device:
(a) \emph{Fast Domain Adaptor} (FDA) aims to learn a global benchmark model based on the history samples of all distributions and generate the network parameters for the distribution-specific device model based on the real-time device samples (in Sec.~\ref{subsec:umn});  (b) \emph{AnchorFrame Distribution Reasoner} (ADR) standardizes the input for the FDA across various multi-model samples. (in Sec.~\ref{subsec:ppg}).

\textbf{Model Pipeline.} As shown in Algorithm \ref{algo:main_algo}, there are three steps in the pipeline of the CDC-MMPG model: (1) Training the cloud model $\mathcal{M}_g(.)$, including the FDA module and the ADR module, with the history samples $\mathcal{S}_{H}$.
(2) Uploading the real-time samples $\mathcal{S}_{R^{(i)}}$ from the device side to the cloud side. Then, the cloud side model $\mathcal{M}_g(.)$ generates the personalized parameters for the device model $\mathcal{M}_{d^(i)}(.)$.
(3) With the model parameters passed from the cloud side, the device model $\mathcal{M}_{d^(i)}(.)$ is updated and makes the final prediction based on the input read-time samples.

\textbf{Multi-modal Feature Extraction.}
We extract the multi-modal representation needed for the subsequent processes from the input examples. 
Specifically, given a real-time input video $ V_{r} $ and a corresponding language query $ Q_{r} $, we employ the DeiT \cite{pmlr-v139-touvron21a} as the visual feature extractor $f_v(.)$ and the BERT-base language model \cite{devlin-etal-2019-bert} as the text encoder $f_t(.)$, to extract the features of the input sample. 
\begin{align}
    F_{v}&=f_v(V_{r}),  \\
    F_{t}&=f_t(Q_{r}),
\end{align}
where $F_{v}=\{F_{v}^i\}_{i=1}^{N_f}$ represents the features of all video frames and ${N_f}$ is the video frame number. $F_{t}$ is the language query feature.


We adopt the spatial-temporal positional embedding and modality type embedding following \cite{wang2022allinone} to the extracted features $F_{v}$ and $F_{t}$ and get the corresponding embedded features $E_{v}=\{E_{v}^i\}_{i=1}^{N_f}$ and $E_{t}$. Then, the cross-modal fusion module $g(.)$ consisted of $t$ transformer layers is employed to fuse the visual features $E_{v}$ and the language feature $E_{t}$:

\begin{equation}
    F_{m}=g(E_{v},E_{t}),
\end{equation}
where $F_{m}=\{F_{m}^i\}_{i=1}^{N_f}$ is the generated multi-modal features after the feature fusion.





\subsection{Fast Domain Adaptor}
\label{subsec:umn}

When adapting the cloud model to the personalized device real-time samples of specific distribution, one straightforward strategy is fine-tuning the cloud model on the device. However, it may be unreachable due to a lack of sufficient training samples and corresponding annotations on the device.  
To tackle the aforementioned challenge, we propose a novel solution called the Fast Domain Adaptor (FDA), which is implemented as a cloud-based service. The FDA is designed to process device-captured images as input and generate customized parameters specifically tailored to the Lightweight Multi-modal Model deployed on the device. These customized parameters are carefully crafted to adapt to the unique data distributions encountered on each individual device.


Specifically, we further sample $ D (D>1) $ frames from $ F_{m} $ randomly, upload the sampled multi-modal features from the device side to the cloud side, and average the included $ D $ frames of multi-modal features into the global representation $F_{g}$. Then, we project the feature $F_{g}$ before it is further analyzed:


\begin{equation}
\label{embedding}
    E_g=f_p(F_{g}),
\end{equation}
where $ f_p(.) $ represents the multi-layer perceptron consisting of two linear layers and a normalization layer. The output feature $ E_g $ is treated as the real-time sample embedding for hyper-network \cite{hyper1,hyper2} input:

\begin{equation}
\label{parameter}
    \Theta_{d}=f_h(E_g),
\end{equation}
where $ f_h(.) $ is the multi-layer perceptron. The generated $ \Theta_{d} $ is the parameters for a single linear layer including both the linear weights and bias, which will be passed to the device model. 


The generated model parameters $ \Theta_{d} $ are passed from the cloud side to the device side. The device model $\mathcal{M}_{d}$ is updated with the model parameters $ \Theta_{d} $ and makes the prediction $ P_{n} $ for the input multi-model real-time sample:
\begin{align}
\label{Hyper}
    P_{n}&=\mathcal{M}_{d}(F_{m};\Theta_{d}),
\end{align}
where $F_{m}$ is the extracted multi-modal feature. 


\subsection{AnchorFrame Distribution Reasoner} \label{subsec:ppg}
\
\newline


The FDA is designed to adapt to diverse multi-modal tasks by utilizing specific data inputs. However, certain tasks, such as Video QA, require transmitting significant amounts of data, such as multiple frames or the entire video, to the FDA. This results in considerable communication costs and bandwidth requirements. To address this challenge and improve the adaptability of the FDA across various multi-modal tasks, we introduce the AnchorFrame Distribution Reasoner (ADR). ADR plays a crucial role in standardizing the input format for the FDA, ensuring compatibility across different multi-modal tasks. 

Specifically, 
during the inference process, we randomly select one frame from the whole video on the device side and upload the multi-modal feature $F^{i}_{m}$ corresponding to the selected $i$-th frame to the cloud side. Then, the \textbf{V}ariational \textbf{A}uto-\textbf{E}ncoder (VAE) including an encoder $ \Psi_{E} $ and a decoder $ \Psi_{D} $ are applied to further analyze.
Firstly, we obtain the latent distribution by analyzing the uploaded feature $F^{i}_{m}$ with the VAE encoder $ \Psi_{E} $:
\begin{equation}
    \Psi_{E}(F^{i}_{m})\sim \mathcal{N}(\mu^{real}, \sigma^{real}),
\end{equation}
where $\mathcal{N}(\mu^{real}, \sigma^{real})$ represents the Gaussian distribution with mean $\mu^{real}$ and standard deviation $\sigma^{real}$.
Then we sample a new representation $ R_{m}^i $ from the obtained specific distribution:

\begin{equation}
    R_{m}^i\sim \mathcal{N}(\mu^{real}, \sigma^{real}).
\end{equation}

The $ R_{m}^i $ is then reasoned on by the VAE decoder $ \Psi_{D} $:

\begin{equation}
    {F^{i}_{m}}^{\prime}=\Psi_{D}(R_{m}^i).
\end{equation}

The $ {F^{i}_{m}}^{\prime} $ and $ F^{i}_{m} $ are utilized to adaptively generate the new feature $F_g'$ through the adaptive generator $\Phi_{a}(.)$. The generated new features will be forwarded to the hyper-network with the goal of leading it to generalize to the target domain. Thus, we can update the Eq.~\ref{embedding} to:

\begin{equation}
    F^{\prime}_{g} = \Phi_{a}({F^{i}_{m}}^{\prime},F^{i}_{m}).
\end{equation}

The feature $F^{\prime}_{g}$ is applied to replace the global representation of the input sample $F_g$ defined in Section ~\ref{subsec:umn} for further prediction of the personalized model parameters.

The framework uses limited communication bandwidth: one-frame multi-modal feature upload and hyper-network parameters (a few linear layers) download, promising the generalization ability and extremely low communication delay of the device model.

During the training process, all history samples are directly saved on the cloud side and participated in the training process, following the setting of previous methods \cite{dcclr}.
Given the sampled multi-modal features $ F_{g} $ and the randomly selected feature $ F_{m}^i $, the variational auto-encoder is applied to further analyze the same as the inference process. Firstly, the encoder $ \Psi_{E} $ project $ F_{m} $ into the latent space and generate a new feature $F_{h}$:

\begin{equation}
    F_{h}=\Psi_{E}(F_{g}).
\end{equation}

Then, to reconstruct the input feature $F_{g}$, the decoder $ \Psi_{d} $ decodes the feature $F_{h}$ to the feature $ F^{\prime}_{g} $:

\begin{equation}
    F^{\prime}_{g}=\Psi_{D}(F_{h}).
\end{equation}

In order to train the variational auto-encoder, we use the \textbf{K}ullback-\textbf{L}eibler (KL) loss $L_{kl}$ and reconstruction loss $L_{Rec}$ as follows:

\begin{align}
    L_{kl}&=\mathbb{D}_{kl}(\mathcal{N}(\mu^{\prime}, \sigma^{\prime})||\mathcal{N}(0,\textit{I})),\\
     L_{\rm{Rec}}&=\rm{MSE}(F_{g},F^{\prime}_{g}),
\end{align}
where $\mathcal{N}(0,\textit{I})$ represents the normal distribution, $\mathbb{D}_{kl}(.)$ represents the Kullback-Leibler divergence calculation and $ \rm{MSE}(\cdot) $ denotes the mean square loss.

\begin{algorithm}[!t]
\SetAlgoLined
{
Initialize the multi-modal feature extraction model: $\mathcal{M}_{vl}$; the hyper-network model in FDA: $ h(\cdot) $; the encoder and the decoder in ADR: $\Psi_{E}, \Psi_{D}$; the adaptive generator: $\Phi_{a}$.\\
\textbf{Phase I:} {Input: Historical data $ \mathcal{S}_{H} $.}\\
\quad 1) Train the ADR on the cloud using $ \mathcal{S}_{H} $ and the training objectives in Eq.~\ref{eq:loss}. \\
Output the trained models: $ \mathcal{M}^{\prime}_{vl}, h^{\prime}(\cdot), \Psi^{\prime}_{E}, \Psi^{\prime}_{D}, \Phi^{\prime}_{a} $. 
\\

\textbf{Phase II:} {Input: Real-time data $ \mathcal{S}_{R} $.} \\
\quad 1) Extract multi-modal features and process the one-frame feature to the cloud $ F^{i}_{cloud}, F_{g} \leftarrow  \mathcal{M}^{\prime}_{vl}(\mathcal{S}_{R})$. \\
\quad 2) Obtain latent distribution of $ F_{cloud} $: $ \mathcal{N}(\mu, \sigma)$, where $\sigma,\mu \leftarrow \Psi^{\prime}_{E}(F^{i}_{cloud})  $. \\ 
\quad 3) Randomly sample from this distribution $F_{rand} \leftarrow \mathcal{N}(\mu, \sigma) $.\\
\quad 4) Decode the randomly sampled distribution $F^{\prime}_{rand} \leftarrow \Psi^{\prime}_{D}(F_{rand}) $.\\
\quad 5) Get adaptive layer embedding $ E_{m} \leftarrow \Phi_{a}(,F^{\prime}_{rand},F^{i}_{cloud}) $.\\
\quad 6) Generate adaptive parameters using the layer embedding $ \Theta_{a} \leftarrow h^{\prime}(E_{m}) $. \\
Output the parameters: $ \Theta_{a} $. 
\\
\textbf{Phase III:} {Input: Real-time data $ \mathcal{S}_{R} $.} \\
\quad 1) Download the adaptive parameters to the device $ \Theta^{d}_{a} \leftarrow \Theta_{a} $.\\
\quad 2) Share feature backbone between the cloud and device $ \mathcal{M}_{d} \leftarrow \mathcal{M}_{vl} $.\\
\quad 3) Make the prediction with $ \Theta^{d}_{a} $: $ P_{n} \leftarrow \mathcal{M}_{d}(\mathcal{S}_{R};\Theta^{d}_{a}) $\\

}
Output the $ P_{n} $.
\caption{\mbox{CDC-MMPG Framework}}
\label{algo:main_algo}
\end{algorithm}

Inspired by \cite{NEURIPS2020_ed265bc9}, in order to incorporate the various distribution shifts into the feature of the single video frame, the adaptive generator generates new features $ F_{a} $ by re-normalizing the $ F^{\prime}_{g} $ to have the same channel-wise mean and standard deviation as the $ F^{i}_{m} $. This process can be formulated as follows: 

\begin{equation}
    \Phi_{a}(F^{\prime}_{g},F^{i}_{m})=\delta (F^{\prime}_{g}) \left (  \frac{F^{i}_{m}-\delta (F^{i}_{m})}{\gamma (F^{i}_{m})} \right ) +\gamma (F^{\prime}_{g}),
\end{equation}
where $ \delta(\cdot) $ and $ \gamma(\cdot) $ denotes channel-wise mean and standard deviation operations, respectively.

Following previous works on variational auto-encoder generalization, we additionally compute the loss values between the generated new features and $ F_{g} $, $ F^{i}_{m} $, respectively, to train the whole module:

\begin{equation}
    L_{\rm{AG}}= \lambda \rm{MSE}(F_{\rm{a}},F_{g})+\rm{MSE}(F_{\rm{a}},F^{i}_{m}),
\end{equation}
where $ \lambda $ denotes a hyper-parameter, and $ F_{\rm{a}} $ denotes the feature generated by the adaptive generator. Thus, the overall training loss $L$ for the whole framework is:

\begin{equation}
    L=L_{\rm{AG}}+L_{\rm{Rec}}+L_{\rm{KL}}.
\label{eq:loss}
\end{equation}

\begin{table}[!h]
    \tabcolsep=0.120cm
    \caption{Results of our proposed method in Open-ended Video QA task on MSRVTT-QA and MSVD-QA. We adopt accuracy as the evaluation metric. The number of learnable parameters on the device, the number of learnable parameters on the cloud, and the time delay are additionally measured to show the efficiency of our proposed method. ``F-linear" denotes only fine-tuning the classifiers after the multi-modal feature extractor. ``F-hyper" denotes only fine-tuning the classifiers and a simple hyper-network without an adaptive generator.}
    \label{TB:opvqa1}
\begin{tabular}{c|cccc|cccc}
\toprule \toprule
\multirow{2}{*}{Methods} & \multicolumn{4}{c|}{VSRVTT-QA}                                                                         & \multicolumn{4}{c}{MVSD-QA}                                                                            \\ \cline{2-9} 
                         & \multicolumn{1}{c|}{Acc.} & \multicolumn{1}{c|}{D-Param.} & \multicolumn{1}{c|}{C-Param.} & Time Delay & \multicolumn{1}{c|}{Acc.} & \multicolumn{1}{c|}{D-Param.} & \multicolumn{1}{c|}{C-Param.} & Time Delay \\ \midrule
F-linear                 & \multicolumn{1}{c|}{13.6} & \multicolumn{1}{c|}{11.6M}    & \multicolumn{1}{c|}{/}        & 60000ms    & \multicolumn{1}{c|}{17.3} & \multicolumn{1}{c|}{11.5M}    & /                             & 60000ms    \\ \midrule
Fine-tuning              & \multicolumn{1}{c|}{36.7} & \multicolumn{1}{c|}{11.6M}    & \multicolumn{1}{c|}{/}        & 60000ms    & \multicolumn{1}{c|}{34.3} & \multicolumn{1}{c|}{11.5M}    & /                             & 60000ms    \\ \midrule
F-hyper                  & \multicolumn{1}{c|}{11.1} & \multicolumn{1}{c|}{11.6M}    & \multicolumn{1}{c|}{28.1M}    & 5.55ms     & \multicolumn{1}{c|}{6.34} & \multicolumn{1}{c|}{11.5M}    & 18.7M                         & 3.71ms     \\ \midrule
Ours                     & \multicolumn{1}{c|}{\textbf{37.1}} & \multicolumn{1}{c|}{11.6M}    & \multicolumn{1}{c|}{55.9M}    & 5.55ms     & \multicolumn{1}{c|}{\textbf{35.4}} & \multicolumn{1}{c|}{11.5M}    & 46.5M                         & 3.71ms     \\ \bottomrule \bottomrule
\end{tabular}
\end{table}

\begin{table}[!h]
    \tabcolsep=0.620cm
    \caption{Results of our proposed method in Open-ended Video QA task on TGIF. We adopt accuracy as the evaluation metric. The number of learnable parameters on the device, the number of learnable parameters on the cloud, and the time delay are additionally measured to show the efficiency of our proposed method. ``F-linear" denotes only fine-tuning the classifiers after the multi-modal feature extractor. ``F-hyper" denotes only fine-tuning the classifiers and a simple hyper-network without an adaptive generator.}
    \label{TB:opvqa2}
\begin{tabular}{c|cccc}
\toprule \toprule
\multirow{2}{*}{Methods} & \multicolumn{4}{c}{TGIF}                                                                                   \\ \cline{2-5} 
                         & \multicolumn{1}{c|}{Accuracy} & \multicolumn{1}{c|}{D-Param.} & \multicolumn{1}{c|}{C-Param.} & Time Delay \\ \midrule
F-linear                 & \multicolumn{1}{c|}{27.3}     & \multicolumn{1}{c|}{11.6M}    & \multicolumn{1}{c|}{/}        & 60000ms    \\ \midrule
Fine-tuning              & \multicolumn{1}{c|}{54.7}     & \multicolumn{1}{c|}{11.6M}    & \multicolumn{1}{c|}{/}        & 60000ms    \\ \midrule
F-hyper                  & \multicolumn{1}{c|}{19.6}     & \multicolumn{1}{c|}{11.6M}    & \multicolumn{1}{c|}{18.7M}    & 5.70ms     \\ \midrule
Ours                     & \multicolumn{1}{c|}{\textbf{55.2}}     & \multicolumn{1}{c|}{11.6M}    & \multicolumn{1}{c|}{46.5M}    & 5.70ms     \\ \bottomrule \bottomrule
\end{tabular}
\end{table}

\section{Experiment}
In this section, we evaluate our proposed method on three popular multi-modal tasks and four relevant datasets. Additionally, we perform ablation studies to investigate the impact of each module in our method, the influence of the number of frames sampled from the whole video, and the time delay between the cloud and the device with various internet conditions.

\subsection{Datasets}
We conducted our experiments on three multi-modal tasks: multiple-choice Video QA, open-ended Video QA, and video-text retrieval. For the open-ended Video QA task, we use the MSRVTT-QA dataset \cite{MSRVTT}, MSVD-QA dataset \cite{MSRVTT} and the TGIF dataset \cite{tgif}. 
For the multiple-choice Video QA task, we use the MSRVTT-QA dataset \cite{MSRVTT}.
For the video-text retrieval task, we experiment on the MSRVTT dataset \cite{xu2016msr-vtt}.

\begin{enumerate}

  \item  \textbf{MSRVTT.}
The MSRVTT is a large-scale dataset for the open-domain video captioning, which consists of 10K video clips and each video clip is annotated with 20 English sentences. The standard splits use 6,513 clips for training, 497 clips for validation, and 2,990 clips for testing.

  \item  \textbf{MSRVTT-QA.}
The MSRVTT-QA dataset contains 10K video clips and 243k question answer pairs. Specifically, we used MSRVTT-QA dataset \cite{MSRVTT} for open-ended VQA and multiple-choice VQA. Following \cite{MSRVTT}, We use the traditional data split which is 65\% for the training set, 5\% for the validation set and 30\% for test set. 

  \item  \textbf{MSVD-QA.}
The MSVD-QA dataset has a total number of 1,970 video clips and 50,505 question answer pairs. Following \cite{MSRVTT}, we split the dataset based on videos that training set takes 61\%, validation set takes 13\%, and test set takes 26\% of the total number of videos.


  \item  \textbf{TGIF.}
The Tumblr GIF (TGIF) dataset contains 100K animated GIFs and 120K sentences describing visual content of the animated GIFs. One sentence is provided for every animated GIF for the training and validation splits, and three sentences per GIF for the test split. Specifically, there are 80K training samples, 11K validation samples and 11k test samples. 

\end{enumerate}

\subsection{Evaluation Metrics}
Following previous works, we adopt accuracy for the open-ended Video QA task, and VR@K, TR@K for video-text retrieval task. Specifically, VR@K denotes the recall of video to text retrieval and TR@K denotes the recall of text to video retrieval. For both of these two tasks, K is set to 1,2,5 respectively. Additionally, we calculate the number of learnable parameters in each model. Meanwhile, for the practical scenario of cloud-device collaboration, we also measure the time delay for the cloud-device communication process.

\begin{table}[!b]
\tabcolsep=0.350cm
\caption{Results of our proposed method in Text-video Retrieval task. The number of learnable parameters on the device, the number of learnable parameters on the cloud, and the time delay are additionally measured to show the efficiency of our proposed method.}
\label{TB:vtr}

\begin{tabular}{cl|cccccc}
\toprule \toprule
\multicolumn{2}{c|}{\multirow{2}{*}{Methods}} & \multicolumn{6}{c}{MSRVTT}                                                                                                                                                                           \\ \cline{3-8} 
\multicolumn{2}{c|}{}                         & \multicolumn{1}{c|}{VR@1}         & \multicolumn{1}{c|}{VR@5}          & \multicolumn{1}{c|}{VR@10}         & \multicolumn{1}{c|}{TR@1}         & \multicolumn{1}{c|}{TR@5}          & TR@10         \\ \midrule
\multicolumn{2}{c|}{F-linear}        & \multicolumn{1}{c|}{2.0}          & \multicolumn{1}{c|}{7.7}           & \multicolumn{1}{c|}{14.1}          & \multicolumn{1}{c|}{3.1}          & \multicolumn{1}{c|}{10.1}          & 16.9          \\ \midrule
\multicolumn{2}{c|}{Fine-tuning}              & \multicolumn{1}{c|}{4.8}          & \multicolumn{1}{c|}{17.6}          & \multicolumn{1}{c|}{27.9}          & \multicolumn{1}{c|}{\textbf{6.2}} & \multicolumn{1}{c|}{20.7}          & 31.8          \\ \midrule
\multicolumn{2}{c|}{F-hyper}             & \multicolumn{1}{c|}{2.6}          & \multicolumn{1}{c|}{8.6}           & \multicolumn{1}{c|}{14.7}          & \multicolumn{1}{c|}{2.7}          & \multicolumn{1}{c|}{11.4}          & 17.6      \\ \midrule     
\multicolumn{2}{c|}{Ours}                     & \multicolumn{1}{c|}{\textbf{6.6}} & \multicolumn{1}{c|}{\textbf{21.3}} & \multicolumn{1}{c|}{\textbf{33.8}} & \multicolumn{1}{c|}{5.6}          & \multicolumn{1}{c|}{\textbf{21.2}} & \textbf{33.3}\\ \bottomrule \bottomrule
\end{tabular}
\end{table}


\begin{table*}[!h]
\tabcolsep=0.250cm
\caption{Ablation studies of different numbers of sampled frames. We make the evaluation on three datasets in Open-ended Video-question Answering task. ``Time/Epoch" denotes the time of fine-tuning for each epoch.}\label{exp:frame}

\begin{tabular}{c|cc|cc|cc}
\toprule \toprule
\multirow{2}{*}{Frames} & \multicolumn{2}{c|}{MSRVTT-QA}                    & \multicolumn{2}{c|}{MSVD-QA}                      & \multicolumn{2}{c}{TGIF}                       \\ \cline{2-7} 
                        & \multicolumn{1}{c|}{Accuracy} & Time/Epoch & \multicolumn{1}{c|}{Accuracy} & Time/Epoch & \multicolumn{1}{c|}{Accuracy} & Time/Epoch \\ \midrule
2                       & \multicolumn{1}{c|}{36.1}         & 740s           & \multicolumn{1}{c|}{34.1}     & 204s           & \multicolumn{1}{c|}{54.3}         & 245s               \\ \midrule
3                       & \multicolumn{1}{c|}{36.7}     & 897s           & \multicolumn{1}{c|}{\textbf{34.3}}     &  228s              & \multicolumn{1}{c|}{54.7}         &  310s          \\ \midrule
4                       & \multicolumn{1}{c|}{37.0}     & 1309s          & \multicolumn{1}{c|}{\textbf{34.3}}     & 318s           & \multicolumn{1}{c|}{54.7}         &   448s             \\ \midrule
5                       & \multicolumn{1}{c|}{\textbf{37.1}}         &  2092s              & \multicolumn{1}{c|}{34.0}     &  377s              & \multicolumn{1}{c|}{\textbf{55.2}}         &   702s         \\ \bottomrule \bottomrule
\end{tabular}
\end{table*}

\subsection{Tasks and Implementation Details} 
We evaluate our method on three tasks: 

\begin{enumerate}
\item \textbf{Video-text Retrieval:} We evaluate two sub-tasks including video-to-text retrieval and text-to-video retrieval. The text-to-video retrieval task requires retrieving the target video using language queries. The video-to-text retrieval task requires retrieving the target text using videos. 

\item \textbf{Open-ended Video-question Answering:} It requires answering questions according to the context of the video. The answers are originally in free-form natural language, but it is a common practice to convert the task to a classification task by representing the answer with a class label.

\item \textbf{Multiple-choice Video-question Answering:} Given a video with a query and 5 candidate captions, the task is to find the one that fits the query out of 5 possible candidates. The correct answer is the ground-truth (GT) caption, and four other negatives are chosen from other captions that have different activity-phrase labels from the correct answer.
\end{enumerate}

 \textbf{Implementation Details.}
We employ the All-in-One-Ti \cite{wang2022allinone} as the baseline model, which includes the DeiT \cite{pmlr-v139-touvron21a} as its visual backbone and BERT-base \cite{devlin-etal-2019-bert} as its semantic encoder. Specifically, we only use the embedding layers of BERT for text embedding. The All-in-one-Ti is adopted as the multi-modal encoder both for the device model and the cloud model. The value of $ D $ is set to 3.
During the training process of ADR, we employ the adamw optimizer with a polynomial decay scheduler. The learning rate is set to 2e-5 and the training epoch is set to 10. During the training process of the rest of the modules, we froze the parameters in the adaptive generator, and the same adamw optimizer is used with the 1e-4 learning rate used. The total training epoch for the rest modules is set to 40. For the hyper-parameters and other settings during the training process, $ \lambda $ is set to 0.1, and the hidden layer of the hyper-network is set to 96 (for Open-ended VQA and Multiple-choice VQA) and 256 (for Video-text Retrieval).
During the inference process, all the parameters on the device model and the cloud model are frozen except the last few dynamic linear layers of the device model. The cloud model and the device model share the same parameters of the multi-modal encoder. Then the cloud generates dynamic parameters using hyper-network for the device model for better generalization.

\subsection{Performance Comparison}
In this section, we introduce how we design the baseline method to showcase the efficiency of our proposed method, and we present the experimental results of our method on various tasks and datasets.

\subsubsection{Baseline Methods.}
\

For we are the first to explore this field, we design the other three baseline methods to show the superiority of our proposed method.
As shown in Tab.~\ref{TB:opvqa1} and Tab.~\ref{TB:opvqa2}, we adopt accuracy as the evaluation metric. The number of learnable parameters on the device, the number of learnable parameters on the cloud, and the time delay are additionally measured to show the efficiency of our proposed method. ``F-linear" denotes only fine-tuning the classifiers after the multi-modal feature extractor. ``F-hyper" denotes only fine-tuning the classifiers and a simple hyper-network without an adaptive generator. Besides our proposed method, we also design the fine-tuning approach, fine-tuning only the linear classifiers approach (F-linear), and fine-tuning both the linear classifiers and solely hyper-network approach. For the fine-tuning approach, we simply add layers of MLPs after the multi-modal feature extractor and do fine-tuning on the whole model. For fine-tuning only the linear classifiers approach, we froze the parameters in the multi-modal feature extractor and fine-tuned the layers of MLPs after the feature extractor. For the fine-tuning of both the linear classifiers and the solely hyper-network approach, we employ a simple hyper-network without the proposed adaptive generator and do fine-tuning for both linear classifiers and the simple hyper-network. 

 \textbf{Open-ended Video-question Answering.}
For Open-ended Video-question Answering, the responses are typically expressed in unrestricted natural language. However, a prevalent approach involves transforming this task into a classification problem by encoding the answers as class labels. To achieve this, we incorporate layers of MLPs with a hidden layer dimension of 96 following the extraction of multi-modal features. The dimension of the MLP's output layer varies according to the specific label size associated with the datasets, for instance, 1501 labels for the MSRVTT-QA dataset.

 \textbf{Multiple-choice Video-question Answering.}
For Multiple-choice Video-question Answering, where both the questions and candidate answers are presented as sentences, our approach involves concatenating the question and the answer candidates. To distinguish between them, we employ the special token [SEP]. Subsequently, we determine the prediction by selecting the candidate with the highest output logit, signifying its likelihood of being the correct answer.

 \textbf{Video-text Retrieval.}
For Video-text Retrieval, the retrieval process encompasses two directions: text-to-video retrieval and video-to-text retrieval. In each direction, the respective modality is first extracted, and subsequently, a comparative analysis is conducted. The predictions are ultimately generated through the application of MLPs.

\subsubsection{Main Results}
\

In this section, we present the experimental results of open-ended Video QA in Tab.~\ref{TB:opvqa1} and Tab.~\ref{TB:opvqa2}, multiple-choice Video QA in Tab.~\ref{TB:muvqa}, and Video-text Retrieval in Tab.~\ref{TB:vtr} on four popular datasets.

\textbf{Effectiveness.} 
Tab.~\ref{TB:opvqa1},Tab.~\ref{TB:opvqa2}, Tab.~\ref{TB:muvqa}, and Tab.~\ref{TB:vtr} provide a comprehensive overview of the superiority of our method compared to other baseline approaches across various datasets and evaluation criteria. Notably, our method consistently outperforms alternative baseline models in nearly all aspects across these datasets.
For example, consider the task of Open-ended Video-question Answering on the MSRVTT-QA dataset. Our proposed method excels in terms of both accuracy and time efficiency when compared to all other baseline models. The conventional fine-tuning method demands a significant amount of time to complete the fine-tuning process. As demonstrated in Tab.~\ref{exp:frame}, when selecting three frames from the entire video, the time required for each fine-tuning epoch on the MSRVTT dataset is approximately 897 seconds. In contrast, our method achieves nearly real-time communication, effectively reducing the time delay between the cloud and the device to as little as 5.55ms, contingent on internet conditions. Furthermore, our proposed method even surpasses the fine-tuning approach in terms of performance, indicating its superior fast generalization capabilities on personalized samples.
Tab.~\ref{TB:opvqa1} and Tab.~\ref{TB:opvqa2}, which denote the results on the MSVD-QA and TGIF datasets, it is evident that our method significantly outperforms other baseline models, thus affirming the effectiveness of our proposed approach.

\textbf{Extensibility.}
Our proposed approach exhibits the capacity to enhance accuracy while concurrently reducing time delays across all datasets associated with the three specific tasks. As delineated in Tab.~\ref{TB:vtr}, our methodology yields a significant performance improvement, particularly in the domain of text-video retrieval. Our assessment encompasses a two-directional approach, addressing both text-to-video retrieval and video-to-text retrieval. Remarkably, our method surpasses the majority of baseline methods, not only in terms of accuracy but also with respect to time delays.
In the context of multiple-choice Video QA, our approach consistently asserts its superiority across all evaluation metrics. This is evident from the data presented in Tab.~\ref{TB:muvqa}.

\begin{table}[]
\tabcolsep=0.60cm
\caption{Results of our proposed CDC-MMPG framework in the multiple-choice Video QA task. The number of learnable parameters on the device, the number of learnable parameters on the cloud, and the time delay are measured to show the efficiency of our proposed method.}

\label{TB:muvqa}

\begin{tabular}{c|ccc|c}
\toprule \toprule
\multirow{2}{*}{Methods} & \multicolumn{3}{c|}{MSRVTT-QA}                                              & \multirow{2}{*}{Time Delay} \\ \cline{2-4}
                         & \multicolumn{1}{c|}{Acc.} & \multicolumn{1}{c|}{D-Param.} & C-Param. &                             \\ \midrule
F-linear           & \multicolumn{1}{c|}{3.58}         & \multicolumn{1}{c|}{11.4M}         &  /        &  $ \ge $ 60000ms                           
 \\ \midrule
Fine-tuning              & \multicolumn{1}{c|}{75.6}         &  \multicolumn{1}{c|}{11.4M}         & /         &  $ \ge $ 60000ms                          \\ \midrule
F-hyper             & \multicolumn{1}{c|}{46.0}         & \multicolumn{1}{c|}{11.4M}         &   37.2K       & $ \ge $ 5.70ms\\ \midrule
Ours                     & \multicolumn{1}{c|}{\textbf{77.0}}         & \multicolumn{1}{c|}{11.4M}         & 20.3M         & $ \ge $ 5.70ms                            
\\ \bottomrule \bottomrule
\end{tabular}
\end{table}

\begin{table*}[t]
\tabcolsep=0.360cm
\caption{Time delay of our framework in different circumstances (various internet speeds) on different datasets. Our proposed method largely reduces the time delay between the cloud and the device, realizing almost real-time communication. $ ``\uparrow" $ denotes the upload process from the device to cloud. $ ``\downarrow" $ denotes the parameters downloaded from the cloud to device.}

\label{TB:internet}

\begin{tabular}{l|l|l|l|l|l}
\toprule \toprule
Datasets    & Size                                                       & 4G: 5MB/s                                                & 4G: 15MB/s                                                & 5G: 50MB/s                                                & 5G: 100MB/s                                                \\ \midrule
MSRVTT      & \begin{tabular}[c]{@{}l@{}}$\uparrow$:0.75KB\\ $\downarrow$:568.5KB\end{tabular} & \begin{tabular}[c]{@{}l@{}}$\uparrow$:0.15ms\\ $\downarrow$:111ms\end{tabular} & \begin{tabular}[c]{@{}l@{}}$\uparrow$:0.05ms\\ $\downarrow$:37.0ms\end{tabular} & \begin{tabular}[c]{@{}l@{}}$\uparrow$:0.01ms\\ $\downarrow$:11.1ms\end{tabular} & \begin{tabular}[c]{@{}l@{}}$\uparrow$:0.007ms\\ $\downarrow$:5.55ms\end{tabular} \\ \midrule
MSVD        & \begin{tabular}[c]{@{}l@{}}$\uparrow$:0.75KB\\ $\downarrow$:379.4KB\end{tabular} & \begin{tabular}[c]{@{}l@{}}$\uparrow$:0.15ms\\ $\downarrow$:74ms\end{tabular}  & \begin{tabular}[c]{@{}l@{}}$\uparrow$:0.05ms\\ $\downarrow$:24.7ms\end{tabular} & \begin{tabular}[c]{@{}l@{}}$\uparrow$:0.01ms\\ $\downarrow$:7.41ms\end{tabular} & \begin{tabular}[c]{@{}l@{}}$\uparrow$:0.007ms\\ $\downarrow$:3.71ms\end{tabular} \\ \midrule
TGIF        & \begin{tabular}[c]{@{}l@{}}$\uparrow$:0.75KB\\ $\downarrow$:583.8KB\end{tabular} & \begin{tabular}[c]{@{}l@{}}$\uparrow$:0.15ms\\ $\downarrow$:114ms\end{tabular} & \begin{tabular}[c]{@{}l@{}}$\uparrow$:0.05ms\\ $\downarrow$:38.0ms\end{tabular} & \begin{tabular}[c]{@{}l@{}}$\uparrow$:0.01ms\\ $\downarrow$:11.4ms\end{tabular} & \begin{tabular}[c]{@{}l@{}}$\uparrow$:0.007ms\\ $\downarrow$:5.70ms\end{tabular}    \\ \bottomrule \bottomrule
\end{tabular}
\end{table*}

\subsection{Ablation Studies}

 \textbf{Number of sampled frames.}
In Tab.~\ref{exp:frame}, we investigate the influence of the number of sampled frames $ D $ from the entire video on the final results. Simultaneously, we measure the time of fine-tuning for each epoch. To be specific, we employ the standard fine-tuning approach to evaluate this factor across three distinct datasets, all in the Open-ended Video-question Answering task. The batch size for all the datasets is set to 256, and the hardware is a single A100 GPU.
It becomes evident that as the number of sampled frames increases, the time of fine-tuning for each epoch experiences a rapid escalation, which is in line with intuition. For instance, when only two frames are sampled, the fine-tuning process demands approximately 740 seconds on MSRVTT-QA dataset. However, if five frames are sampled from each video, the time requirement extends to around 2092 seconds, which is almost three times longer than the former scenario.
Interestingly, the model's performance does not exhibit a consistent upward trend with the increasing number of sampled frames. Besides, the model's performance reaches its peak when the number of sampled frames is set at three or four.
To make a trade-off between performance and training cost, we set the number of sampled frames as three in our model.

 \textbf{Different modules in our proposed CDC-MMPG framework.}
In accordance with the results in Tab.~\ref{TB:opvqa1} and Tab.~\ref{TB:opvqa2}, we have devised three additional baseline methods to assess the efficacy of our comprehensive framework. To be more specific, our assessment involves a comparative analysis of the performance outcomes between the fine-tuning model and our proposed model, serving to emphasize the latter's superiority. Furthermore, we have undertaken an evaluation of the "F-linear" model to facilitate an ablation study of the classifiers subsequent to the multi-modal feature extractor. In addition, the "F-hyper" model has been examined in order to conduct an ablation study focused on the hyper-network and the adaptive generator. It is noteworthy that the omission of our proposed adaptive generator renders a simple hyper-network prone to encountering sub-optimal performance, particularly evident in certain datasets, including MSRVTT-QA, MSVD-QA, and TGIF. This diminished performance is primarily attributed to the issue of over-fitting, leading to the emergence of spurious correlations between visual cues and predictions. Our adaptive generator, as proposed, substantially improves the performance of the hyper-network through the implementation of anchor-frame distribution reasoning.

 \textbf{Time delay between cloud and device in various internet conditions.}
Tab.~\ref{TB:internet} illustrates the time delay associated with data transmission between cloud services and devices under various internet connectivity conditions. Specifically, we have conducted this experiment in four distinct scenarios, each reflecting a different internet condition: 4G (5MB/s), 4G (15MB/s), 5G (50MB/s), and 5G (100MB/s). These scenarios correspond to real-world situations involving older mobile devices with limited 4G connectivity, older mobile devices with optimal 4G connectivity, modern mobile devices with 5G connectivity, and modern mobile devices with advanced 5G internet connectivity.
Our proposed method demonstrates remarkable improvements in terms of time delay. For example, during the inference process, a mere 0.75KB of features is required to be uploaded to the cloud, a task that consumes only about 0.007ms in a 5G (100MB/s) environment. Even in the presence of constrained 4G (5MB/s) internet, the time required for uploading remains negligible. The duration of parameter downloading, however, fluctuates according to the dataset in use, largely due to variations in parameter size. For instance, the parameter size is 583.8KB for the TGIF dataset, the largest among the datasets, whereas the parameter size for the MSVD-QA dataset, the smallest, amounts to 379.4KB. Notably, even when dealing with the demanding TGIF dataset, 
our proposed approach demonstrates the capability to achieve real-time predictions. For instance, in a 5G (100MB/s) environment, the time required for parameter downloading is merely 5.70ms. Even in scenarios featuring limited 4G (5MB/s) connectivity, the time delay remains nearly imperceptible in practical terms.

\section{Conclusion}

 In this paper, we focus on the challenge of enabling efficient and effective adaptation of AI systems to personalized multi-modal data generated by intelligent devices. The shifting data distribution between the cloud and devices necessitates novel approaches to ensure high-quality personalized services. We  introduced a universal on-device Multi-modal Model Adaptation Framework, featuring the Fast Domain Adaptor (FDA) and the AnchorFrame Distribution Reasoner (ADR). FDA, hosted in the cloud, tailors model parameters for on-device Lightweight Multi-modal Models, optimizing adaptability across diverse data distributions. ADR further standardizes input, reducing communication costs. Our contributions, consolidated within the Cloud-Device Collaboration Multi-modal Parameter Generation (CDC-MMPG) framework, constitute a pioneering solution for on-Device Multi-modal Model Adaptation (DMMA), demonstrated through extensive experiments. Looking forward, promising directions include expanding the framework to accommodate various modalities, refining personalized data handling techniques, and further reducing communication costs in multi-modal tasks. 
 


\bibliographystyle{ACM-Reference-Format}
\bibliography{sample-base}

\end{document}